# Multipole Analysis of Radio Continuum Images of Supernova Remnants: Comparison of Type Ia and Core Collapse

**Sujith Ranasinghe\*, Denis Leahy**

Department of Physics & Astronomy, University of Calgary, Calgary, Canada
Email: *syranasi@ucalgary.ca

## Abstract

A multipole expansion analysis is applied to 1420 MHz radio continuum images of supernova remnants (SNRs) in order to compare Type Ia and core collapse (CC) SNRs. Because the radio synchrotron emission is produced at the outer shock between the SNR and the ISM, we are investigating whether the ISM interaction of SNRs is different between Type Ia and CC SNRs. This is in contrast to previous investigations, which have shown that Type Ia and CC SNRs have different asymmetries in the X-ray emission from their ejecta. The sample consists of 19 SNRs which have been classified as either Type Ia or CC. The quadrupole and octupole moments normalized to their monopole moments (total emission) are used as a measure of asymmetry of the emission. A broad range (by a factor of ~1000) is found for both quadrupole and octupole normalized moments. The strongest correlation we find is that large quadrupole moments are associated with large octupole moments, indicating that both serve as similar indicators of asymmetry. The other correlation we find is that both moments increase with SNR age or radius. This indicates that interstellar medium structure is a strong contributor to asymmetries in the radio emission from SNRs. This does not seem to apply to molecular clouds, because we find that association of a SNR with a molecular cloud is not correlated with larger quadrupole or octupole moments.

## 1. Introduction

Supernova remnants (SNRs) are an important source of energy and heavy elements to the interstellar medium (ISM). Determining the progenitor type of a SNR is observationally a difficult task. The progenitor type can be determined by association of a SNR with a neutron star or pulsar wind nebula, or by analysis of the composition of the ejecta of a SNR. But many SNRs do not have sufficiently sensitive X-ray observations for the latter method. For SNRs with absence of the above evidence, attempts have been made to separate Type Ia and core collapse (CC) SNRs by measuring the morphology of SNRs ([1] [2]). The idea is that Type Ia SNRs are more spherical and mirror symmetric than CC SNRs.

In this paper, we investigate the multipole method for analyzing SNR radio images. The method is similar to that of [1], which measured SNR morphology using a multipole expansion analysis of SNR X-ray line emission images. Furthermore, we investigate whether there is a correlation between SNRs with the type of SNR and molecular cloud (MC) associations.

We present the SNR sample and the analysis method in Section 2, and the results of the multipole analysis in Section 3. Implications of the results are given in Section 4.

## 2. Data and Analysis

### 2.1. Data and SNR Selection

We extracted 1420 MHz continuum images of SNRs from the VLA (Very Large Array) Galactic Plane Survey (VGPS: [3]), the Canadian Galactic Plane Survey (CGPS: [4] [5]) and the Southern Galactic Plane Survey (SGPS: [6]). The VGPS 1420 MHz continuum dataset used in the analysis covers a longitude range 18° between and 67°. It has a latitude range between −1.3° and +1.3° for l < 46° and a range between −1.9° and +1.9° for l > 46°. The CGPS dataset covers the region between longitudes 74.2° and 147.3° and between latitudes −1.3° and +5.6°. The SGPS cover range between longitudes 253° and 358° and latitudes −1.5° and +1.5°.

There are ~295 known galactic SNRs [7]. However, there are only a small number of SNRs where the progenitor type is known. Using the SNR catalog (http://www.physics.umanitoba.ca/snr/SNRcat [8]), we found 7 Type Ia SNRs and 20 CC SNRs. We omitted SNRs that were too faint and the ones where the true extent was unclear. The resulting set of SNRs consist of 5 Type Ia and 14 CC SNRs. Table 1 lists the SNRs used in this study.

### 2.2. The Multipole Method

The multipole method was first applied to characterize the X-ray morphology of galaxy clusters by [9]. Reference [10] extended the technique to Chandra observations of SNRs. Furthermore, [2] used the technique to type young SNRs using 24 μm emission data. This method measures the asymmetries in an image by calculating the multipole moments of the surface brightness inside a circular aperture.

The two-dimensional gravitational potential $\Psi(R,\varphi)$ is a solution of the Poisson's equation: $\nabla^2 \Psi(R,\varphi) = 4\pi G \Sigma(R,\varphi)$, where $\nabla$ is the two-dimensional Laplacian, $G$ is the gravitational constant and $\Sigma$ is the surface mass density. For the multipole analysis, the surface mass density $\Sigma(R,\varphi)$ is replaced by the surface brightness [10]. The solution of the Poisson's equation is:

$$\Psi(R,\varphi) = -2Ga_0 \ln\left(\frac{1}{R}\right) - 2G \sum_{m=1}^{\infty} \frac{1}{mR^m}(a_m \cos m\varphi + b_m \sin m\varphi) \quad (1)$$

where the multipole coefficients $a_m$ and $b_m$ are given by:

$$a_m(R) = \int_{R' \leq R} \Sigma(x')(R')^m \cos m\varphi' d^2 x', \quad (2)$$

$$b_m(R) = \int_{R' \leq R} \Sigma(x')(R')^m \sin m\varphi' d^2 x', \quad (3)$$

with $x' = (R', \varphi')$. The multipole powers are given by $P_m(R)$:

$$P_m(R) = \frac{1}{2\pi} \int_0^{2\pi} \Psi_m(R,\varphi) \Psi_m(R,\varphi) d\varphi. \quad (4)$$

For multipole surface brightness analysis, $2G$ is replaced by 1, which reduces the multipole powers to:

$$P_0 = \left[a_0 \ln(R_{ap})\right]^2 \quad (5)$$

Table 1. The coordinates of the centroid and the aperture radius ($R_{ap}$) of Galactic SNRs.

| # | SNR[a] | Centroid Coordinates | | $R_{ap}$ (arcmin) |
|---|---|---|---|---|
| | | $l(°)$ | $b(°)$ | |
| 1 | G41.1-0.3[v] | 41.120 | −0.305 | 4.5 |
| 2 | G43.3-0.2[v] | 43.265 | −0.180 | 4.8 |
| 3 | G120.1+1.4[c] | 120.090 | 1.405 | 4.8 |
| 4 | G337.2-0.7[s] | 337.199 | −0.700 | 4.0 |
| 5 | G352.7-0.1[s] | 352.732 | −0.099 | 4.7 |
| 6 | G21.5-0.9[v] | 21.500 | −0.885 | 2.1 |
| 7 | G27.4+0.0[v] | 27.380 | −0.005 | 3.0 |
| 8 | G29.7-0.3[v] | 29.705 | −0.245 | 3.0 |
| 9 | G31.9+0.0[v] | 31.875 | 0.040 | 4.8 |
| 10 | G33.6+0.1[v] | 33.655 | 0.045 | 7.5 |
| 11 | G39.2-0.3[v] | 39.220 | −0.310 | 4.8 |
| 12 | G109.1-1.0[c] | 109.130 | −1.050 | 18.3 |
| 13 | G116.9+0.2[c] | 116.770 | 0.180 | 35.1 |
| 14 | G130.7+3.1[c] | 130.720 | 3.065 | 5.7 |
| 15 | G332.4-0.4[s] | 332.413 | −0.355 | 6.0 |
| 16 | G337.8-0.1[s] | 337.800 | −0.067 | 4.7 |
| 17 | G348.5+0.1[s] | 348.456 | 0.111 | 6.7 |
| 18 | G348.7+0.3[s] | 348.678 | 0.378 | 5.3 |
| 19 | G349.7+0.2[s] | 349.722 | 0.189 | 2.6 |

[a]The superscripts v, c and s indicate VGPS, CGPS and SGPS surveys respectively.

for $m = 0$ and

$$P_m = \frac{1}{2m^2 R_{ap}^{2m}}\left(a^2 + b^2\right), \tag{6}$$

for $m > 0$. Here $R_{ap}$ is the radius of the circular aperture that contains the object in its entirety.

We divide the powers $P_m$ by $P_0$, to obtain normalized power ratios. The origin is placed at the surface brightness centroid so that $P_1$ vanishes. The higher order terms of the power ratios give information on the morphology of the object. $P_2/P_0$ is the normalized quadruple moment is sensitive to the ellipticity. $P_3/P_0$ is the normalized octupole moment ratio.

Contour plots of the multipole basis functions $r^m \cos m\varphi$ are shown in **Figure 1**. The number of lobes is 2 $m$, with half of them positive and half negative, and higher $m$ values have the lobes more concentrated toward the outer radius. Thus, the multipole analysis is more sensitive to structure near the outer radius than near the center. The corresponding basis functions $r^m \sin m\varphi$ are the same but rotated counter-clockwise by 90˚/m, *i.e.* by half the size of one lobe.

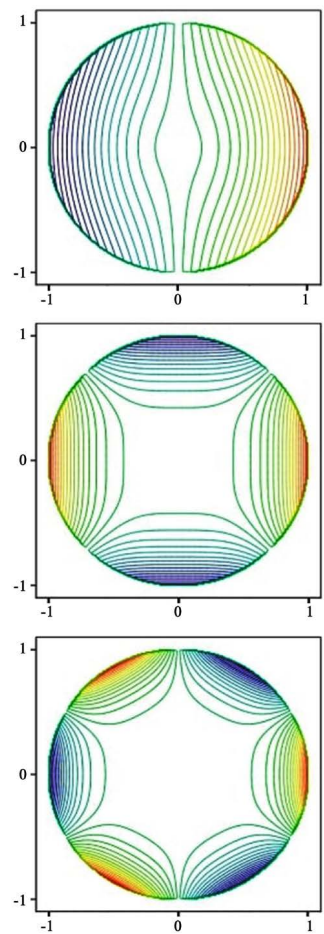

**Figure 1.** Contour plots of the dipole, quadrupole and octupole moments ($m$ = 1, 2 and 3, respectively). Red, green and blue contours represent positive, intermediate and negative values, respectively.

We set a threshold brightness temperature, based on background surface brightness for each SNR, to distinguish it from the background. The center of the circular aperture was set to the centroid of the surface brightness image for each SNR. The aperture radius was determined by choosing the minimum radius which fully enclosed the SNR. Then radius values were scaled by the aperture radius. In Table 1 we list the aperture radii and the centroid coordinates for the SNRs in this study.

### 2.3. Error Analysis

The error estimates for the power ratios were done as follows. We chose several background regions for each SNR image and determine the standard deviation of the background brightness temperature, $\sigma_{T_B}$. We create a noise image with each pixel value chosen randomly from a normal distribution with image mean 0 and with standard deviation $\sigma_{T_B}$. The original image and a noise image are added, then we recalculate the power ratios. This process is repeated 100 times.

The standard deviation of the 100 power ratios was used as the standard error. For SNRs with low surface brightness and high background fluctuations, the error was comparable to the power ratios. We find that the error is roughly equal to the power ratio when the surface brightness of the SNR is low, and the standard deviation of the background brightness temperature is high. (e.g. for G337.2-0.7, we find $P_2/P_0 \approx 10 \pm 10$ (×10$^{-6}$), see Table 2).

Table 2. Galactic supernova remnants and their power ratios.

| # | SNR[a] | Type of SNe | MC[b] Inter: | Distance (kpc) | Age (kyr) | $(P_2/P_0)$ ×10$^6$ | $(P_3/P_0)$ ×10$^6$ | Refs[c] |
|---|---|---|---|---|---|---|---|---|
| 1 | G41.1-0.3[v] | Ia | Yes | 8.5 ± 0.5 | 1.3 | 768 ± 9 | 17.5 ± 0.7 | [11] [12] [13] |
| 2 | G43.3-0.2[v] | Ia | Yes | 11.3 ± 0.4 | 3.3 | 333 ± 3 | 0.74 ± 0.07 | [13] [12] [13] |
| 3 | G120.1+1.4[c] | Ia | Yes? | 4 ± 1 | 0.45 | 13.5 ± 0.4 | 8.31 ± 0.16 | [14] [15], ---- |
| 4 | G337.2-0.7[s] | Ia | No | 2.0 ± 0.5 - 9.3 ± 0.3 | 0.75 - 3.5 | 10 ± 10 | 6.8 ± 5.2 | [16] [17] [17] |
| 5 | G352.7-0.1[s] | Ia | No | 7.6 ± 0.8 | 2.2 | 537 ± 78 | 19.5 ± 8.3 | [16] [18] [19] |
| 6 | G21.5-0.9[v] | CC | No | 4.4 ± 0.2 | 0.47 | 17 ± 1 | 0.036 ± 0.024 | [20] [12] [13] |
| 7 | G27.4+0.0[v] | CC | Yes? | 5.8 ± 0.3 | 2.5 | 242 ± 14 | 4.8 ± 1.1 | [21] [12] [13] |
| 8 | G29.7-0.3[v] | CC | Yes | 5.6 ± 0.3 | 0.89 | 585 ± 7 | 16.3 ± 0.5 | [22] [12] [13] |
| 9 | G31.9+0.0[v] | CC | Yes | 7.1 ± 0.4 | 9 | 53 ± 4 | 14.2 ± 1.0 | [23] [24] [13] |
| 10 | G33.6+0.1[v] | CC | Yes | 3.5 ± 0.3 | 0.78 | 254 ± 22 | 67 ± 4 | [18] [12] [13] |
| 11 | G39.2-0.3[v] | CC | Yes | 8.5 ± 0.5 | 6.2 | 162 ± 12 | 6.1 ± 1.1 | [25] [12] [13] |
| 12 | G109.1-1.0[c] | CC | Yes | 3.1 ± 1.0 | 14 | 8470 ± 60 | 260 ± 6 | [26] [27] [28] |
| 13 | G116.9+0.2[c] | CC | No | 3.1 | 11.5 - 13 | 636 ± 14 | 189 ± 4 | [29] [30] [31] |
| 14 | G130.7+3.1[c] | CC | No | 2.0 | 0.85? | 1084.3 ± 0.7 | 14.38 ± 0.03 | [32] [32] [32] |
| 15 | G332.4-0.4[s] | CC | Yes | 3.1 | 1.35 - 3.05 | 24 ± 10 | 0.67 ± 0.66 | [33] [34] [35] |
| 16 | G337.8-0.1[s] | CC | Yes | 12.3 | 12 - 16 | 920 ± 150 | 24 ± 11 | [36] [37] [38] |
| 17 | G348.5+0.1[s] | CC | Yes | 6.3 - 9.5 | 12 | 440 ± 46 | 8.4 ± 2.6 | [39] [39] [40] |
| 18 | G348.7+0.3[s] | CC | No | 13.2 | 0.35 - 3.15 | 879 ± 73 | 50 ± 12 | [39] [39] [41] |
| 19 | G349.7+0.2[s] | CC | Yes | 11.5 | 1.8 | 40.9 ± 4.2 | 0.15 ± 0.14 | [16] [42] [42] |

[a]The superscripts v, c and s indicate VGPS, CGPS and SGPS surveys respectively. [b]"?" indicates likely/probable molecular interactions. [c]References are for the SNe type, distances and age, respectively.

## 3. Results and Discussion

**Figure 2** shows the 1420 MHz images of the SNRs used for this analysis. The first five images (numbered in red) are Type Ia SNRs and the rest are CC SNRs (numbered in blue). The red and blue dots in each image denote the centroid of each SNR's emission.

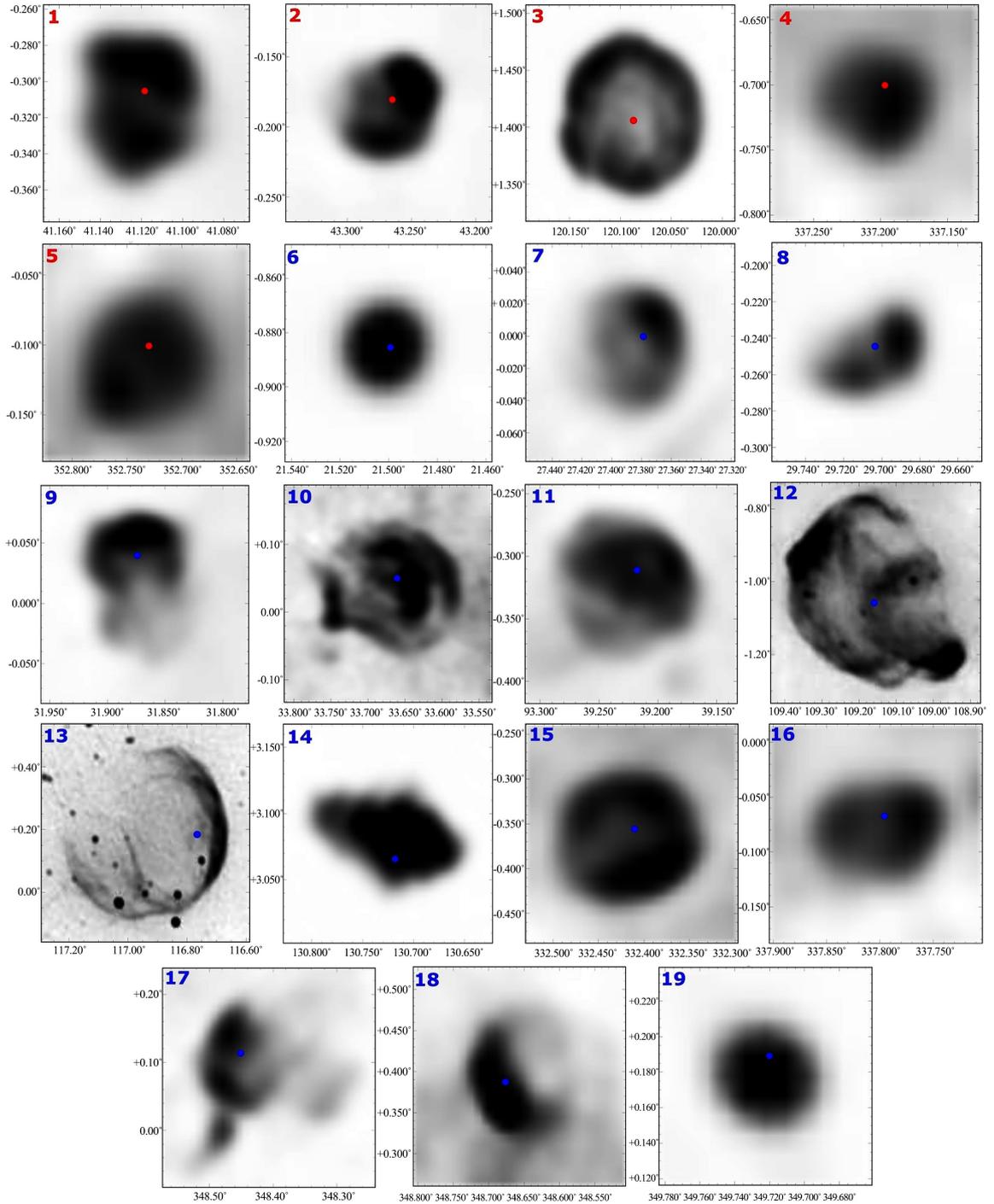

**Figure 2.** 1420 MHz continuum images of SNRs. Red and Blue numbers denote Ia and CC type SNRs respectively and the dots denote the centroid of the SNR.

Table 2 lists the SNRs used in this analysis with their known distances and ages, and our calculated power ratios $P_2/P_0$ and $P_3/P_0$ with errors. Figure 3 is a plot of the power ratios $P_2/P_0$ versus $P_3/P_0$. The star symbols with the black circle denote SNRs that are associated or interacting with molecular clouds, while the solid circles denote SNRs that do not.

The means and standard deviations of the logarithms (to base 10) of power ratios for different subsets of our sample are given in Table 3. We present the values calculated for log (power ratio) rather than power ratio because of the large dynamic range of the power ratios. The arithmetic mean would be dominated by the 2 or 3 SNRs with the largest values, whereas the mean of the logarithms depends on all of the sample values. The means of $\log(P_2/P_0)$ and $\log(P_3/P_0)$ show no significant difference between Type Ia SNRs and CC SNRs. Nor do the means of SNRs with or without molecular cloud associations.

The CC-type SNR G109.1-1.0 (12 in our list) is an outlier: it has the highest quadrupole moment ($P_2/P_0$) in the sample, by factor of 8. It also has the highest octupole ($P_3/P_0$) moment but does not stand out ($P_3/P_0$ is larger by a factor of 1.38). There is a cluster of SNRs consisting of 9 CC-type and 2 Type Ia, which have large quadrupole moments (between ~100 × $10^{-6}$ and ~1000 × $10^{-6}$) and large octupole moments (between ~5 × $10^{-6}$ and ~200 × $10^{-6}$). The 6 SNRs with the lowest $P_2/P_0$ values (<100 × $10^{-6}$) consist of 4 CC-type and 2 Type Ia SNRs. The 4 SNRs with the lowest $P_3/P_0$ values (<1 × $10^{-6}$) consist of 3 CC-type and 1 Type Ia SNRs.

Next, we consider SNRs classified according to whether they have a molecular cloud association or not. The SNRs with molecular cloud association do not show a different distribution of either quadrupole or octupole moments than those without.

The power ratios are plotted versus SNR ages in Figure 4, with ages taken from the literature and listed in Table 2. Ignoring the outlier SNR G109.1-1.0, there is no clear distinction between CC-type and Type Ia SNRs. The average age of the Type Ia SNRs is significantly lower, but that is an artifact of the sample. For older SNRs it is increasingly difficult to identify SNR type by ejecta composition because of the rapidly fading ejecta brightness (e.g. [43]). In contrast for older SNRs, CC-type may be identified by a long-lasting neutron star or pulsar wind nebulae. In Figure 5 we show the power ratios versus SNR radii. The radii are taken as the average of the semi-major and semi-minor axes for SNRs with known distances. The two classes of SNRs are not clearly separated in either quadrupole or octupole moments. The Type Ia SNRs have on average smaller radii, because of the younger ages of Type Ia's in our sample, as noted above.

The Pearson's $r$ values[1] for $P_2/P_0$ versus ages and radii are 0.471 and 0.421, respectively. For $P_3/P_0$, the $r$ values are 0.449 and 0.482, respectively. Thus, the correlation between quadrupole and octupole moments and either age or radius is marginally significant. The best correlation is between $P_2/P_0$ versus $P_3/P_0$ with a Pearson's $r$ value of 0.676.

---

[1]For a sample size of 19, r > 0.456 is significant at the 0.05 level.

Table 3. The means and standard deviations (std) of $\log(P_2/P_0 \times 10^6)$ and $\log(P_3/P_0 \times 10^6)$.

|  | $\log(P_2/P_0 \times 10^6)$ | | $\log(P_3/P_0 \times 10^6)$ | |
| --- | --- | --- | --- | --- |
|  | mean | std | mean | std |
| All SNRs | 2.34 | 0.76 | 0.91 | 0.95 |
| Type Ia | 2.06 | 0.81 | 0.83 | 0.51 |
| CC | 2.44 | 0.72 | 0.93 | 1.06 |
| Associated with MC | 2.36 | 0.73 | 0.88 | 0.87 |
| Not Associated with MC | 2.29 | 0.84 | 0.97 | 1.17 |

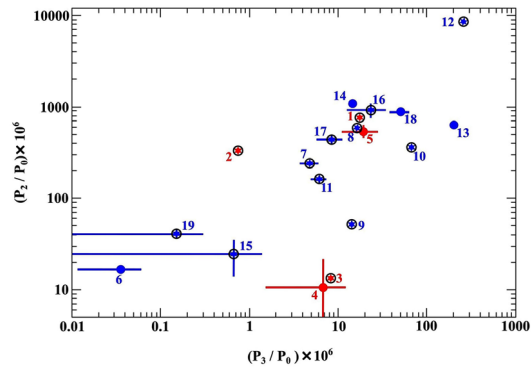

Figure 3. Power ratios, $P_2/P_0$ and $P_3/P_0$ of type-known Galactic SNRs. The red and blue denote Type Ia SNRs and CC SNRs, respectively. The star symbols with black circles denote SNRs with molecular cloud associations.

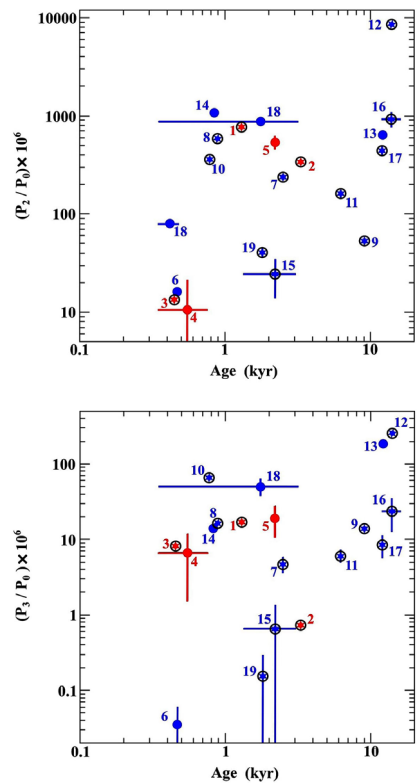

Figure 4. Power ratios versus ages of the type-known SNRs. The red and blue denote Type Ia SNRs and CC SNRs, respectively. The star symbols with black circles denote SNRs with molecular cloud associations.

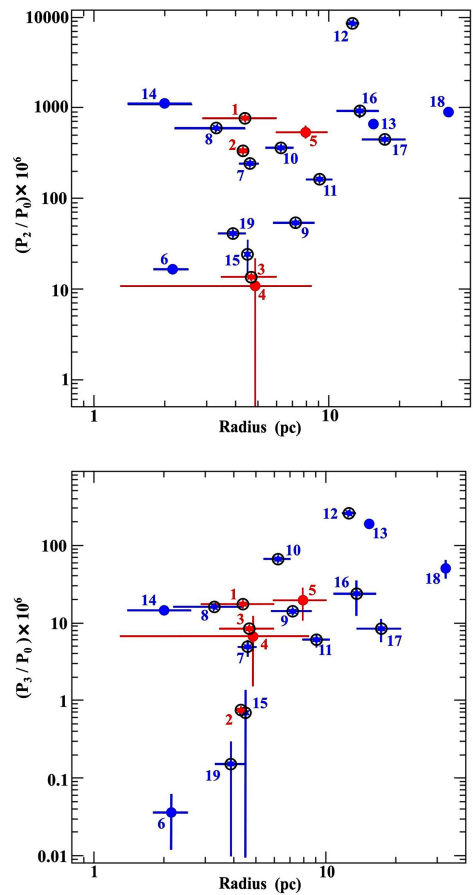

**Figure 5.** Power ratios versus radii of the type-known SNRs. The red and blue denote Type Ia SNRs and CC SNRs, respectively. The star symbols with black circles denote SNRs with molecular cloud associations.

SNR G109.1-1.0 is the most asymmetric SNR of the sample, and SNR G332.4-0.4 is among the set of SNRs with lowest asymmetry (see **Figure 3**). Their images are shown in **Figure 2** (No. 12 and No. 15, respectively). **Figure 6** shows power ratios for these two SNRs as a function of multipole moment, $m$. The difference between G109.1-1.0 and G332.4-0.4 is striking. The power ratios for G109.1-1.0 are significant beyond $m = 14$, whereas G332.4-0.4 has only $m \leq 5$ power ratios which are significant above errors.

## 4. Summary

A multipole analysis was carried out on 1420 MHz radio images for a sample of 19 SNRs of known type (Type Ia and CC). Radio synchrotron emission from SNRs is produced predominantly at the outer shock between the SNR and the ISM. Thus, this study investigates whether the interaction of SNRs with the ISM is different between Type Ia and CC SNRs. The previous investigations of asymmetries in SNRs by [1] and [10] compared the asymmetries in the X-ray emission from the ejecta. They found that Type Ia and CC SNRs have different asymmetries, with CC type showing significantly higher asymmetry.

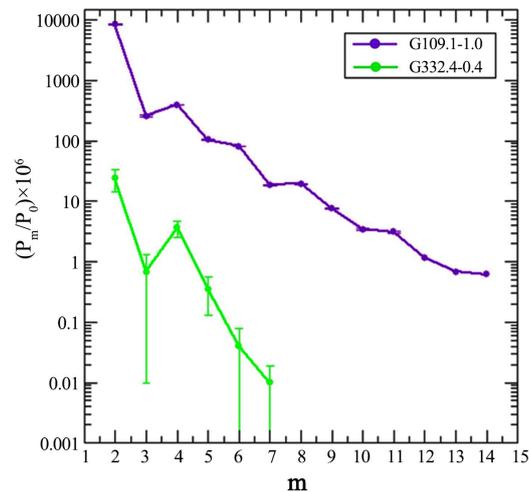

**Figure 6.** Higher orders power ratios of SNRs G109.1-1:0 and G332.4-0.4. Errors on the power ratios are shown by the vertical lines with bar ends.

A broad range (by a factor of 1000) is found here for both quadrupole and octupole normalized moments of the radio emission from SNRs. The strongest correlation that we obtain is that large quadrupole moments are associated with large octupole moments, confirming that these are useful (but different) indicators of asymmetry.

We find that both moments correlate (increase) with SNR age or with SNR radius. We interpret this as structure in the interstellar medium contributing to the asymmetries in the radio emission from SNRs. However, we find that association of a SNR with a molecular cloud is not correlated with larger quadrupole or octupole moments. This seems to imply that molecular clouds do not contribute to asymmetries in SNR radio emission. There may be several reasons that we do not see increased asymmetry for SNRs associated with molecular clouds. Two possibilities are: our sample size is too small; or molecular clouds have lesser effect than other structures in the ISM on radio emission from SNRs.

We find that, solely using the radio data, the two classes of SNRs are not separable due to their morphology/asymmetry as defined by quadrupole and octupole moments. Thus, our conclusion is the ISM interaction of the two classes (Type Ia and CC) of SNRs is not significantly different. This contrasts with the statistically significant difference between the ejecta distributions of Type Ia and CC SNRs [1].

## Acknowledgements

This work was supported by a grant from the Natural Sciences and Engineering Research Council of Canada.

## Conflicts of Interest

The authors declare no conflicts of interest regarding the publication of this paper.